\documentclass[%
reprint,
amsmath,amssymb,
aps,
 prl,
floatfix,
twocolumn]{revtex4-1}

\usepackage{graphicx}
\usepackage{dcolumn}
\usepackage{bm}
\usepackage{amsmath} 
\usepackage{amssymb}
\usepackage{mathrsfs}
\usepackage{siunitx}
\usepackage{xcolor}
\usepackage{tabularx}
\usepackage{microtype}

\newcommand{\ta}{\overline{a}_t}

\definecolor{mytodocolor}{RGB}{200, 87, 0}

\begin{document}

\title{Characterizing spreading dynamics of subsampled systems with non-stationary external input}

\author{Jorge de Heuvel$^1$}
\author{Jens Wilting$^1$}%
\author{Moritz Becker$^{1,2}$}
\author{Viola Priesemann$^1$}%
\thanks{Authors contributed equally}
\author{Johannes Zierenberg$^1$}%
\email{\mbox{johannes.zierenberg@ds.mpg.de}}
\thanks{Authors contributed equally}
\affiliation{$^1$ Max Planck Institute for Dynamics and Self-Organization, G\"ottingen, Germany,}
\affiliation{$^2$ Department of Computational Neuroscience, Third Institute of Physics -- Biophysics, Georg-August-University, G\"ottingen, Germany}

\date{\today}

\begin{abstract}
Many systems with propagation dynamics, such as spike propagation in neural networks and spreading of infectious diseases, can be approximated by autoregressive models. The estimation of model parameters can be complicated by the experimental limitation that one observes only a fraction of the system (subsampling) and potentially time-dependent parameters, leading to incorrect estimates.
We show analytically how to overcome the subsampling bias when estimating the propagation rate for systems with certain non-stationary external input.
This approach is readily applicable to trial-based experimental setups and seasonal fluctuations, as demonstrated on spike recordings from monkey prefrontal cortex and spreading of norovirus and measles.
\end{abstract}

\pacs{Valid PACS appear here}
\maketitle

Propagation dynamics in complex networks are often approximated by models with an autoregressive representation. Examples include affinity maturation in immune systems \cite{nourmohammad_fierce_2019}, reproductive dynamics of bacteria \cite{good_effective_2018, kendall_stochastic_1949, munoz_colloquium_2018, kozlovsky_lubricating_1999} or humans \cite{puerto_branching_2016}, epidemiological disease spreading in a network of humans \cite{farrington_branching_2003, diekmann_definition_1990}, neutron transport theory \cite{pazy_branching_1973} and collective cortical dynamics \cite{beggs_neuronal_2003, haldeman_critical_2005, wilting_inferring_2018, zierenberg_description_2020, neto2019unified, hagemann_no_2020}. The inference of propagation dynamics is often complicated. First, only a fraction of all system components can be observed experimentally (subsampling)~\cite{priesemann_subsampling_2009, ribeiro_spike_2010,  levina_subsampling_2017, wilting_inferring_2018}.  Second, the model parameters can be time-dependent (non-stationary), and specific time-dependent input rates can lead to signatures of criticality even for networks of uncorrelated units~\cite{priesemann_can_2018}. In general, time-dependent input rates are ubiquitous for collective dynamics in neural networks, and are one source for seasonal fluctuations of infectious disease incidence~\cite{franke_discrete-time_2006}.

The subsampling challenge is typically addressed for stationary model parameters. Recent progress has been made for equilibrium and non-equilibrium systems by explicitly modelling the hidden units~\cite{bravi_inferring_2017, dunn_learning_2013, bravi_inference_2017, bachschmid-romano_inferring_2014, dunn_appropriateness_2017, das_systematic_2019}. 
However, explicit knowledge about the hidden units cannot be guaranteed for real-world applications. A subsampling-invariant approach that does not require knowledge about the underlying model size was recently proposed~\cite{wilting_inferring_2018}. The authors showed that established estimators based on linear regression or Kalman filtering underestimate the propagation behaviour. They introduced a novel multistep regression (MR) estimator that is subsampling invariant by characterizing propagation dynamics through the systems autocorrelation time $\tau$. However, it does not consider time-dependent model parameters.

To tackle non-stationarities, recent approaches considered models with time-dependent parameters. 
Examples include Bayesian models based on Cox-processes~\cite{donner_ecient_2018}, weighted least-squares~\cite{rao_fitting_1970}, or expectation-maximization based on Kalman filtering~\cite{ghahramani_parameter_1996, shumway_approach_1982}. However, none of these methods consider the complication of subsampling, although real spreading processes are usually subsampled \cite{wilting_inferring_2018, papoz_case_1996-1}.

In this Letter, we derive an estimator for a subsampled process subject to a specific type of non-stationary external input, namely cyclostationary input. We first show that the subsampling-invariant MR estimator \cite{wilting_inferring_2018} can be biased if the external input rate changes over time. We then analytically derive a generalization of the MR estimator that can overcome the bias in the case of cyclostationary input.
This approach is subsampling invariant and readily applicable to two prevalent situations:
First, to trial-based experiments, which are commonly found in neuroscience;
Second, to periodic input rates, e.g. the seasonal fluctuations of
infectious disease incidence \cite{franke_discrete-time_2006}. We demonstrate the applicability of our methodology on numerical data (testing robustness to relaxation of our assumptions) and on real-world experimental data.

\begin{figure*}[t]
\centering
\includegraphics[width=0.85\textwidth]{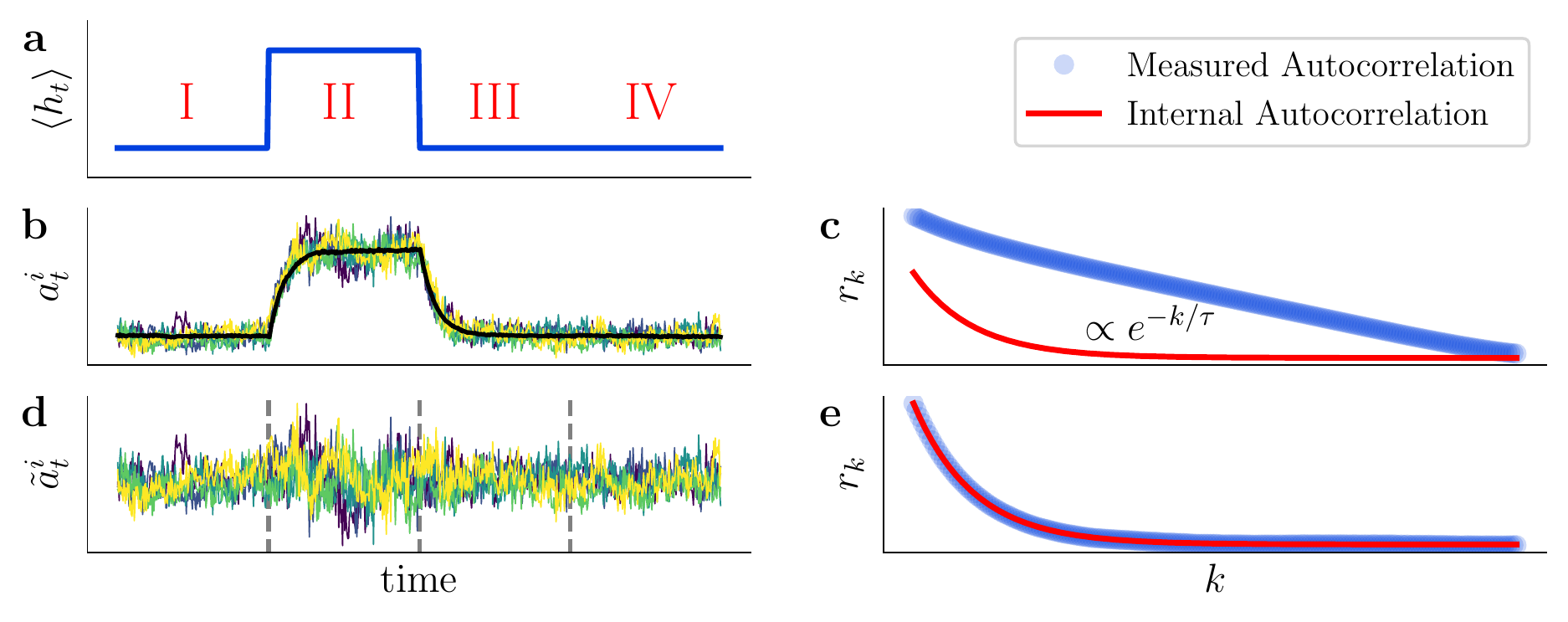}
\caption{
		Unbiased estimation of internal autocorrelation time $\tau$ for a subsampled system with time-dependent input rate can be achieved after subtracting trial-ensemble average from activity.
		\textbf{(a)} Step function as example for a time-dependent input rate. 
		\textbf{(b)} Subsampled activity $a^i_t$ of a branching process with constant internal autocorrelation time $\tau$ and non-stationary input rate shows non-stationary behavior in regime II and III. Colored lines are individual trials, black solid line is trial-ensemble average $\ta$ (over 200 trials).
		\textbf{(c)} Linear regression slope estimates $r_k$ (blue dots) of time lag $k$ for process in (b) do not decay exponentially as expected from the process' internal autocorrelation (red line), which makes an unbiased estimation of $\tau$ impossible.
		\textbf{(d)} Time series (b) corrected by subtracting trial-ensemble average: $\tilde{a}^i_t = a^i_t - \overline{a}_t$.
		\textbf{(e)} For corrected time series (d), the $r_k$ decay exponentially with $\tau$, such that $\tau$ can now be inferred without bias.
		Simulation parameters: 
		Trial length $T=10000$ steps, 
		internal autocorrelation time $\tau=20$ steps, 
		number of trials $N=200$, 
		mean (fully sampled) baseline activity $\left\langle A_0 \right\rangle= 1000$, 
		subsampling fraction $\alpha =0.05$, 
		relative step height $\left\langle h_\mathrm{up} \right\rangle / \left\langle h_\mathrm{down} \right\rangle= 2.6$, 
		step duration $c=200$ steps. 
        \label{Fig:BP_four_regimes_figure_gridplot}
}
\end{figure*}

We consider the class of stochastic processes with an autoregressive representation of first order. This includes widely-used processes, such as branching processes, Kesten processes, and AR(1) processes. Time evolves in discrete steps ($\Delta t = 1$).
Let $A^i_t$ denote the activity of a realization $i$ at time $t$, then the conditional expectation value over the ensemble of independent realizations is defined as 
\begin{equation}
\label{eq:A_t+1}
\left\langle A^i_{t+1} | A^i_t \right\rangle  = m A^i_t + \left\langle h_t \right\rangle,
\end{equation}
where $m$ is the time-independent mean offspring parameter and $\left\langle h_t \right\rangle$ is the average ensemble rate of a time-dependent input distribution. In the framework of spike propagation in neural networks, $m$ describes the average number of neurons that a single neuron subsequently activates and $\left\langle h_t \right\rangle$ describes the expected input rate at time $t$ from sensory modalities or other brain areas. 

Note that the expectation values in Eq.~\eqref{eq:A_t+1} are defined over the ensemble of independent realizations (trials) of the stochastic process, e.g., $\langle h_t\rangle = \sum_{i} h_t^i$ (for the trial-average we drop the index that was summed over). For a general non-stationary external input, $\left\langle h_t \right\rangle$ cannot be defined unless one has multiple realizations from the same time-dependent distributions $h_t^i\sim P(h_t)$. In nature, this is approximately realized by cyclostationarity, e.g., trial-based experiments or seasonal fluctuations. We make use of this to solve the problem even without knowledge of the precise realization of external inputs. In the following, we assume that the generation of offsprings is Poisson distributed with time-independent $m$, while the generation of external input is Poisson distributed with time-dependent rate $\left\langle h_t \right\rangle.$

Subsampling is incorporated as follows: We only require that the subsampled activity $a_t$ is on average linear in the full activity $A^i_t$, i.e., $\langle a^i_t | A^i_t\rangle = \alpha A^i_t$ (for details see Ref.~\cite{wilting_inferring_2018}). For example, every spike or disease incidence is sampled with probability $p=\alpha$. 

To estimate the spreading behaviour $m$ under subsampling and time-dependent external input rates, we follow the principle idea of the MR estimator~\cite{wilting_inferring_2018}. We generalize Eq.~(\ref{eq:A_t+1}) by recursive iteration to $k$ time steps:
\begin{equation}
\label{eq:A_t+k_non-stationary}
\left\langle A^i_{t+k} | A^i_t \right\rangle  = m^k A^i_t +  \sum_{l=1}^{k} m^{k-l} \cdot \left\langle h_{t+l-1} \right\rangle
\end{equation}

If the rate is time-independent ($\left\langle h_t \right\rangle = \left\langle h \right\rangle$),   Eq.~(\ref{eq:A_t+k_non-stationary}) implies that the original process $A_t^i$ has an exponential autocorrelation function 
\begin{equation}
\label{eq:autocorrelation}
	C(k)=m^k=\exp(-k\Delta t/\tau)\text{,}
\end{equation}
with the time lag $k$ in steps of $\Delta t$. 
The autocorrelation function relates the propagation dynamics ($m$) to an internal autocorrelation time $\tau = -\Delta t/\ln(m)$ and represents a measure of how long information persists in the activity~\cite{wilting_inferring_2018}. For stationary processes the variance across trials is equal to the variance within trials ($\text{Var}_i(a^i_t)=\text{Var}_t(a^i_t)=\text{Var}_{i,t}(a_t^i)$), such that the autocorrelation function $C(k)$ of the subsampled activity $a_t^i$ can be calculated directly via linear regression~\cite{wilting_inferring_2018} 
\begin{equation}
\label{eq:regression_slopes}
C(k) = \hat{r}_k = \frac{\text{Cov}_{i,t}(a^i_t, a^i_{t+k})}{\text{Var}_{i,t}(a^i_t)} = \alpha^2  \frac{\text{Var}_{i,t}(A^i_t)}{\text{Var}_{i,t}(a^i_t)} m^k \text{,}
\end{equation}
with time-independent autocorrelation strength $b = \alpha^2 \text{Var}_{i,t}(A^i_t)/\text{Var}_{i,t}(a^i_t)$ for all $k\neq0$. While $b$ is biased under subsampling ($b<1$ if $\alpha < 1$), the autocorrelation time $\tau$ is subsampling invariant and can be obtained by fitting Eq.~\eqref{eq:regression_slopes} to the data~\cite{wilting_inferring_2018}.

For a time-dependent external input rate $\left\langle h_t \right\rangle$, however, the autocorrelation function is not time invariant, and if calculated does not necessarily decay exponentially (Fig.~\ref{Fig:BP_four_regimes_figure_gridplot}a-c). Consider, for example, a step-function external input rate. Linear regression applied to each regime independently would yield similar slopes (identical slopes for full activity $A_t^i$) but different offsets of linear regression (Supplemental Material, Fig.~S2). Therefore, the naive application of the MR estimator fails even for full activity. This represents an issue for general time-dependent input. 

In the following, we construct a reliable estimate of the internal autocorrelation time $\tau$ in the presence of cyclostationary external input rates.
We focus our discussion on subsampled activity $a_t$, which includes the fully-sampled case ($\alpha=1$, $b_t=1$).

To correct the bias from cyclostationary external input ($\left\langle h_t \right\rangle$ is time-dependent but identical for each trial $i$), we introduce the following method: Given we have $N$ trials, with independent realizations $a^i$ of a subsampled linear autoregressive process, we calculate the time-dependent trial-ensemble average
\begin{equation}
\overline{a}_t = \frac{1}{N}\sum_{i=1}^{N} a_{t}^i
\end{equation}
over all trials (not to be confused with an average calculated over all recorded times). Now, we correct for the non-stationarity of the original process by subtracting the trial-ensemble average (Fig.~\ref{Fig:BP_four_regimes_figure_gridplot}d)
\begin{equation}
\tilde{a}_t^i = a_t^i - \overline{a}_t \text{.}
\end{equation}
Its linear regression slopes $r_k$ reveal the true internal autocorrelation time in their exponential decay (Fig.~\ref{Fig:BP_four_regimes_figure_gridplot}e) for sufficiently large $N$ (see below and Supplemental Material Fig.~S4). 

\begin{figure}[t]
	\includegraphics[width=1\linewidth]{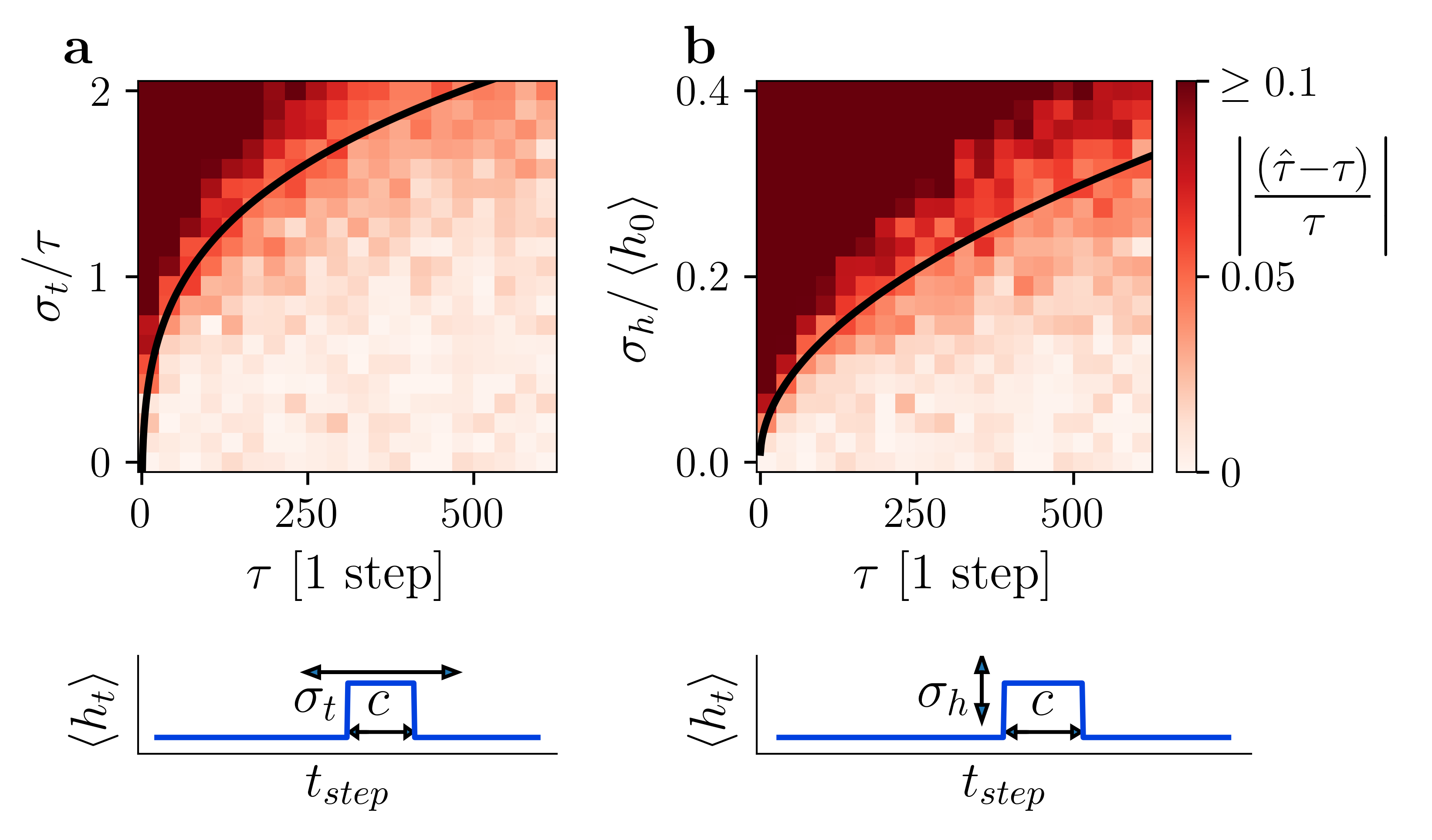}
	\caption{
		Robustness of our estimate to variability in a non-stationary input (step function with step-duration $c=10\text{ }\tau$). 
		\textbf{(a)} Variability in the onset time $t_\mathrm{step}$ with standard deviation $\sigma_t$.
		\textbf{(b)} Variability in the step height $\Delta \left\langle h \right\rangle$ with standard deviation $\sigma_h$. 
		Color in matrix indicates relative error between the estimated autocorrelation time $\hat{\tau}$ and the internal autocorrelation time $\tau$ of the branching process. The \SI{5}{\percent} error bound was fitted (black lines) and scales as $\sigma \propto \tau^\gamma$.
		Simulation parameters: 
	    $T=1000\text{ }\tau$ steps, 
		$N=300$, 
		$\left\langle A_0 \right\rangle= 5000$, 
		$\alpha =0.01$.
		\label{Fig:Estimator_matrix_substracted_sigma}
	}
\end{figure}

From the corrected time series $\tilde{a}_t^i$, we can thus infer the unbiased autocorrelation time by applying the MR estimator~\cite{spitzner_toolbox_2018} (see Supplemental Material S.6 for the full derivation). To prove this, we reformulate Eq.~\eqref{eq:regression_slopes} as simple linear regression at each time across trials, i.e., $\hat{r}_{k,t}=\text{Cov}_i(\tilde{a}_t^i,\tilde{a}_{t+k}^i)/\text{Var}_i(\tilde{a}_t^i)$. For  trial-ensemble corrected $\tilde{a}_t^i$, we find that the correction compensates the convolution in Eq.~\eqref{eq:A_t+k_non-stationary}, such that $\hat{r}_{k,t} = b_t m^k$ with time-independent decay but with time-dependent autocorrelation strength (Supplemental Material, Eq.~(S23))
\begin{equation}
\label{eq:b_t}
b_t = \alpha^2 \frac{\text{Var}_i(A^i_t)}{\text{Var}_i(a^i_t)} \approx \frac{1}{1-(1-\alpha^{-1})F_t^{-1}},
\end{equation}
where the relation to the (across-trial) Fano factor of the full activity $F_t=\text{Var}_i(A^i_t)/\left\langle A_t \right\rangle$ is strictly true only for binomial subsampling. 
However, we can show (Supplemental Material, Eq.~(S25)-(S29)) that for the corrected time series direct application of Eq.~\eqref{eq:regression_slopes} with the standard regression approaches yields an unbiased estimate of the internal autocorrelation time $\tau$ despite cyclostationary input and subsampling (for a proof of concept see Fig.~S3). 

In addition to the bias from subsampling or non-stationary input, there can be a bias from short trial length $T$~\cite{marriott_bias_1954} and from small trial number $N$. The short-trial bias can be avoided by estimating both covariance and variance as fluctuations around a global stationary mean (cf. ``stationarymean'' method in Ref~\cite{spitzner_toolbox_2018} with a detailed discussion). For all our analyses (experimental and numerical), we thus use the MR estimator toolbox~\cite{spitzner_toolbox_2018} with ``stationarymean'' method. In principle, this allows for an unbiased estimation down to $N=10$ short trials (Fig.~S4), while of course the variance of the results increases with decreasing $N$ (Supplemental Material, Sec.~S.4 and S.7).

We tested the applicability of MR estimation for cyclostationary external input by increasing the level of realism for a numerical problem. The test case is a baseline rate $\left\langle h_0 \right\rangle$ plus step-function at onset time $t_\mathrm{step}$ with step height $\Delta \left\langle h \right\rangle$ and step duration $c$. We consider three cases:
i) perfect cyclostationarity across trials (Fig.~\ref{Fig:BP_four_regimes_figure_gridplot} and Fig.~S5 for an extreme example),
ii) variation of onset time $t_\mathrm{step}\sim\mathcal{N}\left(T/2,\sigma_t\right)$ with $\Delta \left\langle h \right\rangle=\left\langle h_0 \right\rangle$ fixed (Fig.~\ref{Fig:Estimator_matrix_substracted_sigma}a), and 
iii) variation of the step height $\Delta \left\langle h \right\rangle \sim\mathcal{N}\left(\left\langle h_0 \right\rangle,\sigma_h\right)$ with $t_\mathrm{step}=T/2$ fixed (Fig.~\ref{Fig:Estimator_matrix_substracted_sigma}b). 
We generated $N=200$ trials of branching processes with internal autocorrelation time $\tau$, trial duration $T=\num{1000}\text{ }\tau$, and baseline activity $\left\langle A_0\right\rangle = 5000$ such that $\left\langle h_0 \right\rangle=(1-m)\left\langle A_0 \right\rangle$ ($m=\exp(-\Delta t/\tau)$, $\Delta t = 1~\text{step}$).
This setup allows us to independently investigate variability in onset time and height of the input. 

\begin{figure*}[t]
\centering
\includegraphics[width=1\textwidth]{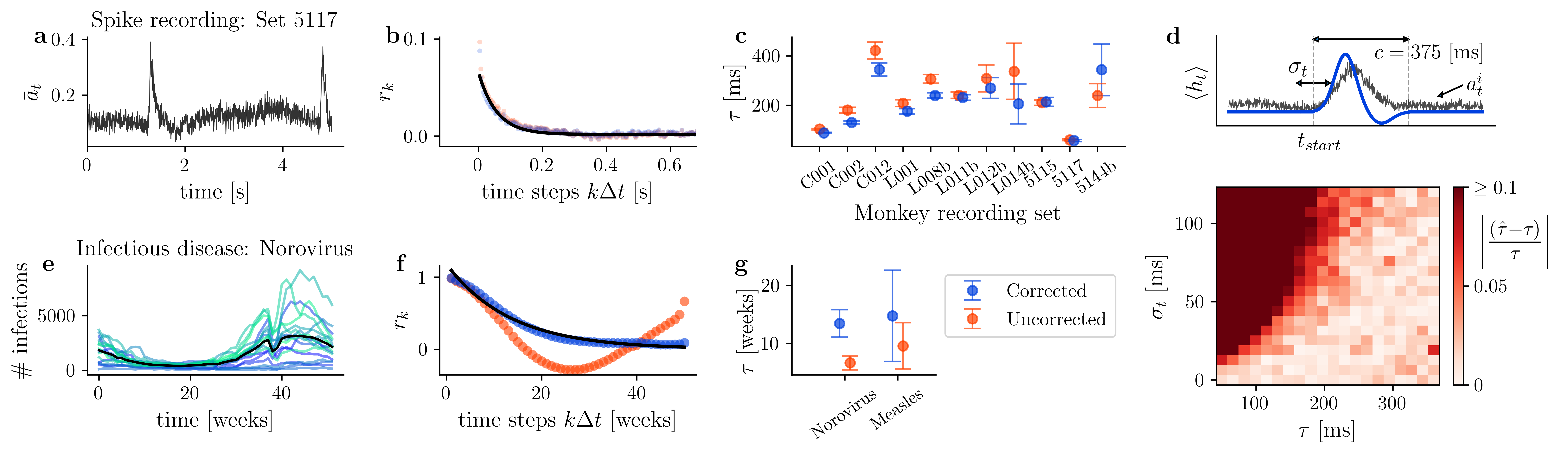}
\caption{
    Application of our new approach to experimental data. 
    \textbf{Top \textbf{(a-c)}:} The intrinsic timescales $\tau$ in macaque pre-frontal cortex have been inferred with our new approach from spike recordings during a trial-based visual short-term memory task~\cite{pipa_performance-_2009}. 
    \textbf{(a)} Example trial-ensemble average of stimulus-evoked non-stationary neural responses.
    \textbf{(b)} Autocorrelation functions $r_k$ of (a) before (orange) and after the correction (blue) hardly differ.
    \textbf{(c)} Intrinsic timescales inferred from uncorrected data are systematically but not very strongly overestimated (less than $10\%$).
    \textbf{(d)} {Numerical robustness validation for typical experimental recordings, resembling a typical evoked potential:
    Non-stationary external input (blue) with 2-fold increase during stimulus presentation (total duration of $c = \SI{375}{\milli\second}$) for $N=300$ trials of length $T=\SI{5}{\second}$ with sampling frequency $f=\SI{1}{\kilo\hertz}$ and subsampling fraction $\alpha=0.01$. An example trial realization $a^i_t$ is shown as black line.
    The impact on estimating $\tau$ under variance of the stimulus onset ($\sigma_t$) is evaluated for various intrinsic autocorrelation times $\tau$, as in Fig.~\ref{Fig:Estimator_matrix_substracted_sigma}.}
    \textbf{Bottom (e-g):} The infectious spreading dynamics of norovirus and measles have been inferred with case report data from the Robert-Koch-Institute~\cite{noauthor_survstatrki_nodate}.
    \textbf{(e)} Reported infection numbers (blue lines) and the time-dependent trial-ensemble average (black line) for norovirus reveal seasonal non-stationarities.
    \textbf{(f)} With our method, the seasonality was mostly removed from the autocorrelation function $r_k$ of (d).
    \textbf{(g)} In contrast to the neural recordings, the infectious spreading dynamics inferred from the uncorrected disease data are systematically underestimated.
	\label{Fig:experimental_results}
}
\end{figure*}

Variations in the onset time and step height do not hinder correct inference as long as the standard deviations are sufficiently low (Fig.~\ref{Fig:Estimator_matrix_substracted_sigma}). In our test case, variations in the onset time barely affect the correct inference as long as the standard deviation $\sigma_t$ is below the magnitude of the autocorrelation time (Fig.~\ref{Fig:Estimator_matrix_substracted_sigma}a). When $\sigma_t \approx \mathcal{O}(\tau)$, the method still provides consistent estimates of the processes autocorrelation time. Moreover, the estimates improve for a given $\sigma/\tau$ with increasing autocorrelation time $\tau$. We observe, that the \SI{5}{\percent} error bound scales as $\sigma_t \propto \tau^\gamma$ with $\hat{\gamma} \approx \num{0.22 \pm 0.03}$. Similarly, variations in the step height barely affect the correct inference as long as the standard deviation $\sigma_h$ is below $\langle h_0\rangle/5$ (Fig.~\ref{Fig:Estimator_matrix_substracted_sigma}b). 
Again, the estimates improve with increasing autocorrelation time and the \SI{5}{\percent} error bound scales as $\sigma_h/\Delta \left\langle h \right\rangle \propto \tau^\gamma$ with $\hat{\gamma} \approx \num{0.4 \pm 0.1}$. We conclude that our method provides consistent results even after relaxation of perfect cyclostationarity.

We applied our method to two sets of experimental data. 
The first dataset consists of spiking activity in pre-frontal cortex from a trial based short-term visual memory task on \textit{macaque mulatta}~\cite{pipa_performance-_2009} (about $N=300$ trials each, see Supplemental Material Sec.~S.11). 
In this dataset, the external input can be interpreted as sensory input from other brain areas to the investigated area. 
The second dataset are epidemiological case reports from the Robert-Koch-Institute~\cite{noauthor_survstatrki_nodate} ($N=18$ trials each, see Supplemental Material Sec.~S.10). 
In the epidemiological dataset, the infections carried into the country via travel can be interpreted as non-stationary external input.

For the monkey data, we want to emphasize three findings: 
First, although the trial-ensemble average $\overline{a}_t$ increases by a factor 3 (Fig.~\ref{Fig:experimental_results}a) the autocorrelation function hardly differs in most cases (Fig.~\ref{Fig:experimental_results}b). 
Second, we find a systematic decrease of intrinsic timescales after correction, while for the majority of the recording sets the decrease was less than $10\%$ (Fig.~\ref{Fig:experimental_results}c). 
Third, a robustness test of our method with parameters adjusted to experimental scale (Fig.~\ref{Fig:experimental_results}d with experimentally realistic stimulus shape) indicates that our method yields less than $5\%$ deviation from $\tau\ge\SI{200}{\milli\second}$ despite stimulus onset variability with $\sigma_t<\SI{50}{\milli\second}$, which is a realistic constraint given the steep rise of typical ensemble responses within \SIrange[tophrase={--}]{30}{50}{ms} (Fig.~\ref{Fig:experimental_results}a). 
To conclude, our method reveals intrinsic timescales in pre-frontal cortex between $\SI{57 \pm 4}{\milli\second}$ and $\SI{345 \pm 26}{\milli\second}$ with median $\SI{214}{\milli\second}$ (compared to $\SI{239}{\milli\second}$ if not corrected) from recordings covering the full task. Our results are consistent with previous results in pre-frontal areas of macaque (about $\SI{200}{\milli\second}$) confined to the stimulus foreperiod to approximate the resting state~\cite{murray_hierarchy_2014, chaudhuri2015large}. 

In the example of disease spreading, our method accounts well for seasonal fluctuations (Fig.~\ref{Fig:experimental_results}e-g). 
The weekly case number reports reveal a strong yearly periodicity, suggesting a year-wise separation into trials. 
The improvement due to trial-ensemble average correction is readily visible in the regression function $r_k$ (Fig.~\ref{Fig:experimental_results}f).
With the correction, the infectiousness estimate is higher than without (Fig.~\ref{Fig:experimental_results}g, Norovirus: $\tau=\SI{14(3)}{weeks}$, Measle: $\tau=\SI{15(8)}{weeks}$). The disease results are in principle subject to additional uncertainty from the small number of trials (cf. Fig.~S4), which are probably on the order of $10\%$ and thus smaller than the error bars from the fits.
Our results highlight that the correction by trial-ensemble average can reveal higher infectiousness of diseases, which might otherwise be underestimated due to seasonal fluctuations and other non-stationary effects, and that long-term recordings are necessary to reveal the intrinsic infectiousness of a disease.

In summary, we have presented a simple, subsampling-invariant estimate of the internal autocorrelation time for stochastic processes with an autoregressive representation subject to (approximate) cyclostationary external input. The key success of the presented approach (MR estimation with trial-ensemble average corrected time series) is the potential to disentangle the internal spreading from any hidden, but repetitive external input rate. Thereby, our approach solves the problem of apparent criticality due to non-stationary input rates~\cite{priesemann_can_2018} for repetitive stimulation protocols. We demonstrated the robustness of our approach to violations of perfect cyclostationarity for the external input rate; and we showed its applicability to real-world problems from neuroscience and epidemiology. In conclusion, we recommend the trial-ensemble average correction as best practise when approximating trial-based experiments with autoregressive models. A toolbox for the multistep-regression analysis is readily available~\cite{spitzner_toolbox_2018}.

\begin{acknowledgments}
We thank Matthias Munk for sharing his data. All authors acknowledge support by the Max Planck Society. 
Financial support was received from the Gertrud-Reemtsma-Stiftung (JW), the Joachim Herz Stiftung (JZ), and the German Ministry of Education and Research (BMBF) via the Bernstein Center for Computational Neuroscience (BCCN) G{\"o}ttingen under Grant No.~01GQ1005B (MB, VP, JZ). 
\end{acknowledgments}

\bibliographystyle{apsrev4-2}
\bibliography{./literatur}

\newpage

\setcounter{equation}{0}
\setcounter{figure}{0}
\setcounter{table}{0}
\setcounter{section}{1}
\setcounter{subsection}{0}

\renewcommand{\theequation}{S\arabic{equation}}
\renewcommand{\thefigure}{S\arabic{figure}}
\renewcommand{\thetable}{S\arabic{table}}
\renewcommand{\thesubsection}{S.\arabic{subsection}}

\section{Supplemental Material}

\subsection{S.1 List of notations}
\begin{table}[!ht]
	\centering
	\begin{tabularx}{\linewidth}{lX}
		Notation & Description \\
		\hline
		$A_t^i$ &  Fully sampled activity of trial $i$ of the autoregressive process at time step $t$ \\
		$a_t^i$ & Subsampled activity of trial $i$ of the autoregressive process at time step $t$ \\
		$\langle A_t\rangle$ & Expectation value over independent realizations for a given time (dropped $i$)\\
		$\left[ A^i\right]$ & Expectation value over time for a given realization (dropped $t$)\\		$m$ & Branching/offspring parameter\\
		$H_t$ & External input at time step $t$ \\
		$\left\langle h_t \right\rangle$  & Mean external input rate at time step $t$ \\
		$T$ & Total length of a time series\\
		$N$ & Number of trials \\		
		$\tau$ & Autocorrelation time \\
		$\Delta t$ & Time step length (absolute)\\
		$k$ & Relative time lag \\
		$b$ & Subsampling induced correlation bias\\
		$\alpha$ & Subsampling fraction\\
		$F$ & Fano factor\\
		$\hat{r}_k$ & Linear regression slope estimate for time lag $k$\\
		$\hat{s}_k$ & Linear regression offset estimate for time lag $k$\\

		$\text{Var}_i(\cdot)$ & Variance over index $i$ \\
		$\text{Cov}_i(\cdot)$ & Covariance over index $i$\\
		\hline
	\end{tabularx}
	\caption{List of notations.
		\label{tab:notations}}
\end{table}

\subsection{S.2 Branching process}
\label{sec:supp_branching_process}
The branching process (BP) with immigration is a stochastic autoregressive process. Each realization of the process $i$ is described by a temporal evolution with time $t$. For realization $i$ at time $t$ there are $A_t^i$ units, of which each unit $j$ generates a random integer number of ``offsprings'' $y_{i,t,j}$ and all $y_{i,t,j} \in \mathbb{N}$ are independently and identically distributed with the mean $m$~\cite{pakes_branching_1971, harris_theory_1963, heathcote_branching_1965}.  Additionally, a time-dependent external input $H^i_t$, with mean $\left\langle h_t \right\rangle$ ``immigrates'' at each time step, where $\langle\cdot\rangle$ denotes the expectation value over independent realizations. The evolution of the total activity $A^i_t$ of the branching process is recursively given by
\begin{equation}
A^i_{t+1}  = \sum_{j=1}^{A^i_t} y_{i,t,j} +H^i_t \text{.}
\end{equation}

\paragraph{\textbf{Autocorrelation time}}
The branching parameter $m$ is directly connected to the processes autocorrelation time~$\tau$ by~\cite{wilting_inferring_2018} 
\begin{equation}\label{eqSupplm-tau}
m = \exp (-\Delta t/\tau) \text{,}
\end{equation}
given a time binning $\Delta t$, e.g., from simulation steps or data binning in experiments.

\paragraph{\textbf{Stationary BP}}
\label{par:supp_stationary_BP}
Assuming the mean of the branching parameter $m$ and the external input $\left\langle h_t \right\rangle$ are constant over time, i.e., $\left\langle h_t \right\rangle=\langle h\rangle$, we can derive the dynamics of the branching process. First, the expectation value of the time step $A^i_{t+1}$ given the activity $A^i_t$ [cf. Eq.(1)] becomes
\begin{equation}
\label{eq:A_t+1_supp}
\left\langle A^i_{t+1} | A^i_t \right\rangle  = m A^i_t + \left\langle h \right\rangle\text{.}
\end{equation}
Recursive iteration of Eq.~(\ref{eq:A_t+1_supp}) and identification of the geometric series yields the expectation value of the evolution over $k$ time steps
\begin{equation}
\label{eq:A_t+k_supp}
\left\langle A^i_{t+k} | A^i_t \right\rangle  = m^k A^i_t +\left\langle h\right\rangle\frac{1-m^k}{1-m}\text{.}
\end{equation}
Now, we can separate the dynamics into three regimes: Subcritical for $m<1$, critical for $m=1$ and supercritical for $m>1$. We find a stationary solution $\left\langle A_\infty \right\rangle $ for the subcritical case by iterating Eq.~(\ref{eq:A_t+k_supp}) to $k\longrightarrow \infty$ ($m^k\to 0$), such that
\begin{equation}
\label{eq:BP_infty_supp}
\left\langle A_{\infty} \right\rangle  = \frac{\left\langle h\right\rangle}{1-m}\text{.}
\end{equation}
In the critical state, the mean activity shows linear growth due to $\left\langle h\right\rangle$, whereas the activity diverges exponentially in the supercritical state. \\

\paragraph{\textbf{BP with non-stationary input}}
\label{par:supp_non-stationary_BP_supp}
Since our main manuscript addresses non-stationary external input, we here investigate a non-stationary branching process with a time-dependent external input $\left\langle h_t \right\rangle$. The stationary distribution in Eq. (\ref{eq:BP_infty_supp}) is no longer valid and by iterating Eq.~(\ref{eq:A_t+1_supp}) with the time-dependent external input rate $\left\langle h_t \right\rangle$  we can derive the conditional expectation value of the activity after $k$ time steps:
\begin{equation}
\label{eq:A_t+k_non-stationary_spp}
\left\langle A^i_{t+k} | A^i_t \right\rangle  = m^k A^i_t +  \sum_{l=1}^{k} m^{k-l} \cdot \left\langle h_{t+l-1}\right\rangle\text{.}
\end{equation}

\subsection{S.3 Subsampling}
\label{sec:supp_subsampling}
When only a fraction of the full system can be observed, this is defined as subsampling. Examples include electrophysiological recordings of neuronal activity in neuroscience or incomplete case reporting of infectious disease propagation.
Naive analysis of the data neglecting the influence of subsampling can lead to severe misinterpretations of the system's dynamics~\cite{priesemann_subsampling_2009, levina_subsampling_2017, wilting_inferring_2018}.\\
The theory and implications of subsampling for linear autoregressive processes have been described in detail in Ref.~\cite{wilting_inferring_2018} and will here be recapitulated briefly. The time series $a_t$ is called a subsample of $A^i_t$, if 
\begin{equation}
\label{eq:subsampling_supp}
\left\langle a^i_t | A^i_t \right\rangle = \alpha A^i_t  + \beta
\end{equation}
holds for all $t$, $j \in \mathbb{N}$ with constants $\alpha$, $\beta \in \mathbb{R}$. The subsample $a^i_t$ is constructed from the fully sampled time series upon sampling and does not interfere with it's evolution. We assume $\beta =0$.\\

\subsection{S.4 Multistep regression estimation}
\label{sec:supp_mre}
To infer a network's autocorrelation time $\tau$ and the branching parameter $m$ even under subsampling, Wilting \& Priesemann developed the multistep regression (MR) estimator \cite{wilting_inferring_2018}. It addresses the issue of classical estimators being biased under subsampling. The MR estimator is applicable to stationary autoregressive processes of first order only, giving misestimations when applied to a non-stationary autoregressive time series. \\
The MR estimator works as follows: In a first step, we estimate the linear correlation of Eq.~(\ref{eq:A_t+k_supp}) between a step $a^i_t$ and $a^i_{t+k}$ (within the same realization) with the slope $r_k$ and offset $s_k$ for time lags $0<k<k_{max}$ for the time steps $t<T-k$, by minimizing the sum of residuals
\begin{equation}
\label{eq:step1_supp}
R_k(\hat{r}_k,\hat{s}_k) = \sum_i^N \sum_{t}^{T-k} (a^i_{t+k} - (\hat{r}_k \cdot a^i_t + \hat{s}_k))^2 \text{.}
\end{equation}
It can be shown~\cite{wilting_inferring_2018}, that for stationary dynamics $\hat{r}_k$ converges in probability to
\begin{equation} 
\label{eq:r_k_and_s_k}
\hat{r}_k \longrightarrow b m^k \text{,}
\end{equation}
where $m^k$ is the slope between the fully sampled activity pairs and $b$ the bias in the slope estimation due to subsampling. More specifically, the linear regression slopes fulfill the relation 
\begin{equation}
\hat{r}_k =  \alpha^2\frac{\text{Var}_i(A^i_t)}{\text{Var}_i(a^i_t)} \text{ } m^k =b  m^k \text{,}
\end{equation}
where the notation $\text{Var}_i(\cdot)$ denotes the variance over independent realizations.
The bias depends on the subsampling fraction $\alpha$ (see Eq.~\ref{eq:subsampling_supp}), the variance of the full activity $A^i_t$ and the subsampled activity $a^i_t$ respectively. However, these are usually unknown. 
Then, in the second step of the estimator, the sum of residuals is minimized for
\begin{equation}
\label{eq:step2_supp}
R(\hat{b},\hat{m}) = \sum_{k} (\hat{r}_k - \hat{b} \cdot \hat{m}^k)^2 \text{,}
\end{equation}
where the two step estimation over various time lags $k$ allows us to infer the bias $b$, which remains unknown to classical linear-regression estimators, and the branching parameter $m$. The autocorrelation time $\tau$ can easily be calculated via Eq.~\eqref{eqSupplm-tau}. The procedure is equivalent to the calculation of time series autocorrelation that has a decreased correlation strength $b$ in step 1 and fitting the exponential decay in step 2.

We used for all analyses the python toolbox \textit{Mr. Estimator}~\cite{spitzner_toolbox_2018} of the multistep regression estimator. The exponential function $f_{\text{exp}}(x,\tau,b) = b \text{ } \exp(-x/\tau)$ has been used for the purpose of this investigation as it addresses the pure exponential decay characteristic for the autocorrelation function of an autoregressive process. For the monkey dataset, the offset-exponential fit-function $f_{\text{exp}}(x,\tau,b,c) = b \text{ } \exp(-x/\tau) + c$ has been used.

Two estimation methods are implemented in the \textit{Mr. Estimator} toolbox. The method \textit{stationarymean} uses all trials combined to calculate the activity average $\overline{a}$, which is needed to calculate the linear regression slopes $\hat{r}_k$ numerically. The advantage of the method \textit{stationarymean} is that the linear regression estimation is more robust if only short trials with few datapoints each are available. 
The method \textit{trialseparated} calculates the activity average $\overline{a}^i$ and subsequently linear regression slopes for each trial $i$ independently and averages over all obtained regression slopes $\hat{r}_{k,i}$, see Ref.~\cite{spitzner_toolbox_2018} for further details. In case the mean activity between trials varies significantly, the method \textit{trialseparated} provides better estimation results for the regression slopes. However, when each trial is short but activity across trials is stationary (or as in our case cyclostationary) the method \textit{stationarymean} corrects for short-trial biases~\cite{spitzner_toolbox_2018}. We validate the trial ensemble average correction on both methods in Sec.~S.8. For all estimations in the paper, the method \textit{stationarymean} was used to correct for short-trial biases.

\subsection{S.5 Effect of non-stationary input on MR estimation}
Assuming a branching process that is subject to a time-dependent external input rate $\left\langle h_t \right\rangle$, a naive application of the MR estimator gives a biased estimation $\hat{\tau}$. We will demonstrate this analytically in the following example of a step function, which can be generalized to arbitrary time-dependent external input rates.

Let $\left\lbrace A^i_t\right\rbrace_{t=0}^T$ be subject to a time-dependent external input rate $\left\langle h_t \right\rangle$ with a step function:

\begin{align}
\label{eq:h_step}
\left\langle h_t \right\rangle = \begin{cases}
h_1 \hspace{1em} \forall  t<t_{step} \\
h_2 \hspace{1em} \forall  t \geq t_{step}
\end{cases}
\end{align}

\begin{figure}[t]
\centering
\includegraphics[width=1\linewidth]{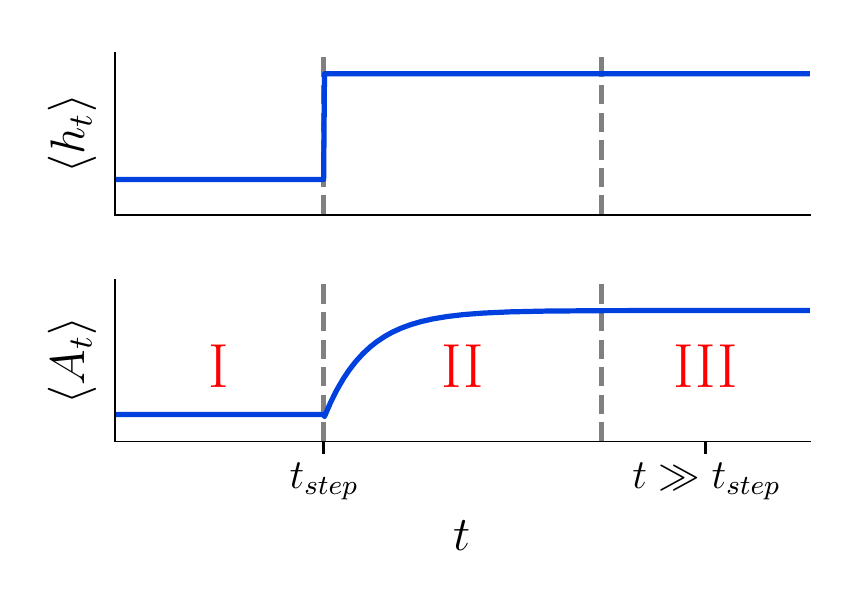}
	\caption{
		Mean activity $\left\langle A_{t} \right\rangle$ (bottom) of a BP subject to a non-stationary input $\left\langle h_{t} \right\rangle$ (top) with step-function rate [Eq.~\eqref{eq:h_step}] can be divided in three regimes. Mean baseline activity $\left\langle A_0 \right\rangle$ in regime I, the transient regime II with growing activity and a new mean activity regime III, compare Eq.~\eqref{eq:A_step}.
		\label{fig:A_mean_regimes}}
\end{figure}

Now, one can divide the mean activity development into three regimes, as shown in Fig.~\ref{fig:A_mean_regimes}. Two stationary regimes I and III with different expectation values and one transient regime II right after the jump in the external input rate. The expectation values for the regimes can be derived from Eqs.~(\ref{eq:BP_infty_supp}) and (\ref{eq:A_t+k_supp}).

\begin{align}
\label{eq:A_step}
\left\langle A_t \right\rangle  \approx  \begin{cases}
\frac{h_1 }{1-m} &\forall  t<t_{step}\\
m^{t-t_{step}} \frac{h_1}{1-m} +h_2\frac{1-m^{t-t_{step}}}{1-m} &\forall  t \geq t_{step}\\
\frac{h_2}{1-m} &\forall t \gg t_{step}
\end{cases}
\end{align}
Here, case 1 ($t<t_{step}$) and 3 ($t \gg t_{step}$) follow immediately from Eq.~(\ref{eq:BP_infty_supp}), while for case 2 ($t \geq t_{step}$) we assume the stationary solution of regime I ($\frac{h_1}{1-m}$) at $t_{step}$, insert this into Eq.~(\ref{eq:A_t+k_supp}) with $\langle h\rangle = h_2$, and identify $k=t-t_{step}$.

By applying the estimator to the different regimes separately, following the steps in Ref.~\cite{wilting_inferring_2018}, one finds that the linear regression offset estimator $\hat{s}_k$ will take on different values due to $\left\langle h_t \right\rangle$  in the different regimes:

\begin{align}
\label{eq:s_k_step}
\hat{s}_{k,I} &\longrightarrow h_1\frac{1-m^k}{1-m} \text{ in regime I,}\nonumber\\
\hat{s}_{k,II} &\longrightarrow h_2\frac{1-m^k}{1-m} \text{ in regime II,}\\
\hat{s}_{k,III} &\longrightarrow h_2\frac{1-m^k}{1-m} \text{ in regime III.}\nonumber
\end{align}

\begin{figure}[t]
	\centering
	\includegraphics[width=1\linewidth]{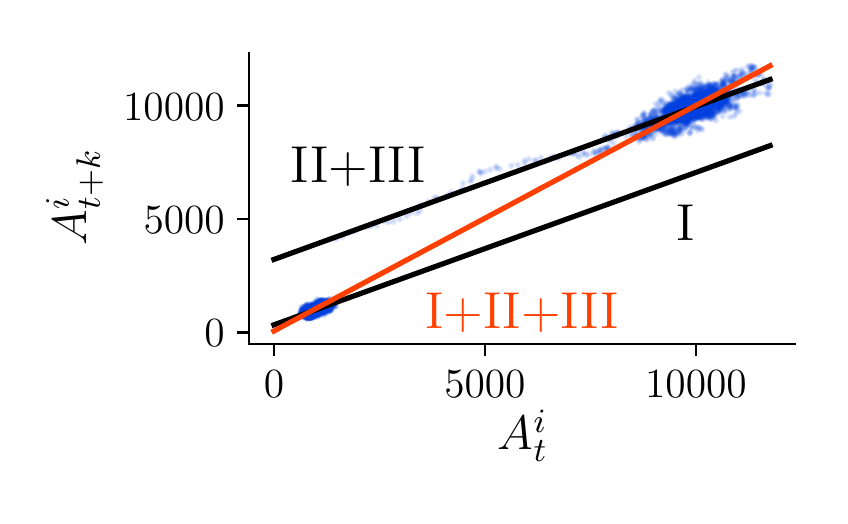}
	\caption{
		Least square estimation of $\hat{r}_k$ [Eq.~(\ref{eq:step1_supp})] for BP subject to a non-stationary input with step-function rate [Eq.~\eqref{eq:h_step}]. The lower left  cloud results from data pairs $(A_t^i, A_{t+k}^i)$ for $t<t_{step}$ (I), the upper right from those for $t \gg t_{step}$ (III). The colored lines represent fits of regimes I and II+III (black lines, no bias in slope) or I+II+III (red line, strong bias in slope), cf. Eq.~(\ref{eq:s_k_step}).
    		Simulation parameters: 
        	Trial length $T=100 \text{ }\tau$ steps, 
        	internal autocorrelation time $\tau=50$ steps,
    		number of trials $N=1$, 
    		mean baseline activity $\left\langle A_0 \right\rangle= 1000$ in regime I, 
    		subsampling fraction $\alpha =1$, 
    		relative step height $\left\langle h_\mathrm{up} \right\rangle / \left\langle h_\mathrm{down} \right\rangle= 10$.
		\label{fig:data_cloud}}
	
\end{figure}

Here we can clearly see that the least square estimation with Eq.~(\ref{eq:step1_supp}) on the entire time series is influenced by the step function in $\left\langle h_t \right\rangle$. As actually two different offsets would be treated as values of unity. Consequently, the estimation of $\hat{r}_k$ will be biased by the time-dependence in the external input rate. To visualize that analytical example, the linear regression for a given time lag $k$ for the regimes I to III separately and combined is visualized in Fig.~\ref{fig:data_cloud}. 

\subsection{S.6 Analytic derivation that linear regression on trial-ensemble average corrected time series allows to infer spreading dynamics}
\label{sec:supp_trial_structure}
This section addresses the trial-ensemble average correction of non-stationary autoregressive time series of first order, to realize a correct estimation of the processes spreading dynamics in terms of the branching parameter $m$ and autocorrelation time $\tau$ respectively and proves the validity analytically.
We discuss subsampled systems with $\left\langle a^i_t | A^i_t \right\rangle = \alpha A^i_t$.
Any following results are applicable to fully sampled systems $A^i_t$ by choosing $\alpha = 1$. 

When the external input $H_t$ is drawn from the same time-dependent probability distribution, we can define a trial ensemble average
\begin{equation}
\overline{a}_t = \frac{1}{N}\sum_{i=1}^{N} a_{t}^i \text{,}
\end{equation}
that averages over all trials for each time step and where $N$ is the number of trials. For $N\longrightarrow \infty$ the trial ensemble average converges in probability to the time-dependent expectation value, thus $\overline{a}_t \longrightarrow  \ta $ at a given time step, which we assume for the following analytical derivations.

We start by defining the mean-corrected time series 
\begin{equation}\label{eq:mean_corr_supp}
\tilde{a}^i_t = a^i_t - \langle a_t\rangle
\end{equation}
such that $\left\langle \tilde{a}_t\right\rangle = 0$. We recall that the actual time evolution takes place in the original process $A^i_t$.

Next, we show that the slopes $r_{k,t}$ from linear regressions over mean-corrected subsampled activities from an ensemble of trials at time $t$ and time $t+k$ can be decomposed into a time-dependent correlation bias $b_t$ and a $k$-dependent decay $m^k$. For each time $t$ we can solve the simple linear regression, Eq.~\eqref{eq:step1_supp}, with
\begin{equation}
r_{k,t} = \frac{\text{Cov}_i(\tilde{a}^i_{t}, \tilde{a}^i_{t+k})}{\text{Var}_i(\tilde{a
}^i_t)}\text{,}
\end{equation}
where the covariance is given by
\begin{equation}
\text{Cov}_i(\tilde{a}^i_{t}, \tilde{a}^i_{t+k}) = \left\langle \tilde{a}_{t}  \tilde{a}_{t+k}\right\rangle - \left\langle \tilde{a}_{t} \right\rangle \left\langle  \tilde{a}_{t+k} \right\rangle \text{,}
\label{eq:covariance_supp}
\end{equation}
and $\langle\cdot\rangle$ denotes the ensemble expectation value (where by convention the index $i$ is dropped). Per construction $\langle\tilde{a}_{t}\rangle=\langle\tilde{a}_{t+k}\rangle=0$, cf. Eq.~\eqref{eq:mean_corr_supp}.
We thus only need to calculate
\begin{align}
\left\langle \tilde{a}_{t}\tilde{a}_{t+k}\right\rangle 
&= \left\langle (a_t-\langle a_t\rangle)(a_{t+k} - \langle a_{t+k}\rangle\right\rangle\\
&= \langle a_t a_{t+k}\rangle - \langle a_t\rangle\langle a_{t+k}\rangle\\
&= \alpha^2\left(\langle A_tA_{t+k}\rangle - \langle A_t\rangle\langle A_{t+k}\rangle\right)
\end{align}
where we used $\langle a_t\rangle = \langle\langle a_t|A_t\rangle\rangle$ and Eq.~\eqref{eq:subsampling_supp}. Using again the law of total expectation, namely $\langle A_tA_{t+k}\rangle=\langle\langle A_tA_{t+k}|A_t\rangle\rangle$ and $\langle A_{t+k}\rangle=\langle\langle A_{t+k}|A_t\rangle\rangle$, we find
\begin{align}
\left\langle \tilde{a}_{t}\tilde{a}_{t+k}\right\rangle 
&= \alpha^2\left(\langle A_t\langle A_{t+k}|A_t\rangle\rangle - \langle A_t\rangle\langle\langle A_{t+k}|A_t\rangle\rangle\right)\nonumber\\
&= \alpha^2\left(m^k\langle A_t^2\rangle + \langle A_t\rangle \sum_{l=1}^{k} m^{k-l} \cdot \left\langle h_{t+l-1}\right\rangle\right. \nonumber\\
&\phantom{=(} \left.- m^k\langle A_t\rangle^2 - \langle A_t\rangle \sum_{l=1}^{k} m^{k-l} \cdot \left\langle h_{t+l-1}\right\rangle\right),\nonumber\\
&=\alpha^2m^k \text{Var}_i(A^i_t).\label{eq:cov_supp}
\end{align}
where we used Eq.~\eqref{eq:A_t+k_supp}. We thus find
\begin{equation}\label{eq:rkt_supp}
    r_{k,t} = \alpha^2\frac{\text{Var}_i(A^i_t)}{\text{Var}_i(\tilde{a^i_t})}m^k = b_t m^k\text{,}
\end{equation}
with a time-dependent amplitude (or bias) $b_t=\alpha^2{\text{Var}_i(A^i_t)}/{\text{Var}_i(\tilde{a^i_t})}$ and a purely $k$-dependent decay $m^k$.

The time-dependent amplitude can be related to the Fano-factor of the original process. To see this, we note that per construction the trial-ensemble expectation value $\langle{\tilde{a}_t}\rangle=0$, such that $\text{Var}_i(\tilde{a}^i_t)=\langle \tilde{a}_t^2\rangle=\text{Var}_i(a^i_t)$. When the subsampling procedure can be described by a binomial distribution, where $\text{Var}_i(a^i_t|A^i_t)=\alpha(1-\alpha)A^i_t$, we obtain from the law of total variance, $\text{Var}_i(a^i_t)=\langle \text{Var}_i(a^i_t|A^i_t)\rangle + \text{Var}_i(\langle a^i_t | A^i_t\rangle) = \alpha(1-\alpha)A^i_t + \alpha\text{Var}_i(A^i_t)$. With the Fano factor $F_t=\text{Var}_i(A^i_t)/\left\langle A^i_t \right\rangle$, the amplitude thus becomes
\begin{equation}
\label{eq:b_t_supp}
b_t = \alpha^2 \frac{\text{Var}_i(A^i_t)}{\text{Var}_i(a^i_t)} = \frac{1}{1-(1-\alpha^{-1})F_t^{-1}}.
\end{equation}

Finally, we show that the time-dependent amplitude $b_t$ still allows the application of the linear regression estimator $\hat{r}_k$ to the mean-corrected subsampled process as found in Eq.~(\ref{eq:step1_supp}) and Eq.~(\ref{eq:step2_supp}) despite cyclostationary external input. For this, it is important to notice that the minimization in the simple linear regression step, Eq.~(\ref{eq:step1_supp}), is solved by 
\begin{equation}\label{eq:rk_nonstat}
    r_k = \frac{\text{Cov}_{i,t}(\tilde{a}_t^i, \tilde{a}_{t+k}^i)}{\text{Var}_{i,t}(\tilde{a}_t^i)},
\end{equation}
where both covariance and variance here run over trial ensemble ($i$) as well as time ($t$).

\begin{figure*}[t]
	\centering
	\includegraphics[width=0.8\textwidth]{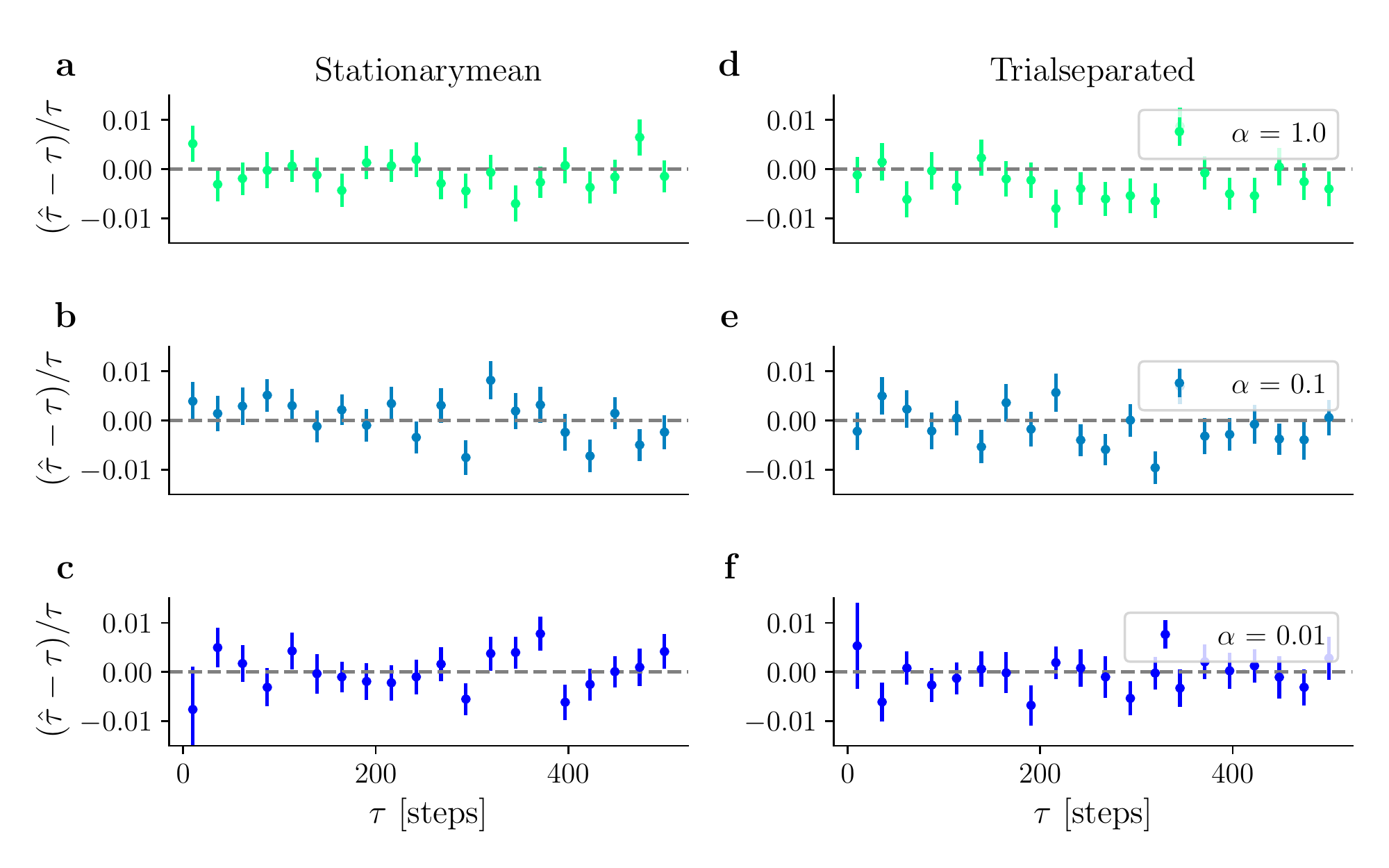}
	\caption{
		Proof of concept A. Numerical validation of the trial-ensemble average correction for subsampled BP with time-dependent external input rate (step function as in Fig. S1) for various autocorrelation times $\tau$ and subsampling fractions $\alpha$. Error bars denote bootstrap error from a single MR estimation with the MR toolbox~\cite{spitzner_toolbox_2018}.
		\textbf{(a-c)} Method \textit{stationarymean}. 
		(Simulation parameters: $T=100 \text{ } \tau$, $N=5000$, $\left\langle A_0 \right\rangle= 5 000$, $\left\langle h_\mathrm{up} \right\rangle / \left\langle h_\mathrm{down} \right\rangle = 2$.)
		\textbf{(d-f)} Method \textit{trialseparated}.
		(Simulation parameters: as before, but $T=5000 \text{ } \tau$ to avoid a short-trial bias for this test [cf. Fig. S4].)
		All estimates lie in the range of \SI{1}{\percent} relative deviation for all simulated autocorrelation times $\tau$ and subsampling fractions $\alpha$.
		\label{fig:poc}
	}
\end{figure*}

For the \textit{stationarymean} method of the MR estimator~\cite{spitzner_toolbox_2018}, Eq.~\eqref{eq:rk_nonstat} translates to
\begin{equation}\label{eq:bias_sm_pre}
    r_k = \frac{\frac{1}{T}\sum_t \langle (\tilde{a}_t -\langle \tilde{a}\rangle)(\tilde{a}_{t+k} -\langle \tilde{a}\rangle)\rangle}{\frac{1}{T}\sum_t \langle (\tilde{a}_t -\langle \tilde{a}\rangle)^2\rangle},
\end{equation}
where $\langle \tilde{a}\rangle=\frac{1}{TN}\sum_{t,i}\tilde{a}_t^i = 0$ by construction. We thus find that
\begin{equation}\label{eq:bias_sm}
    r_k = \frac{\frac{1}{T}\sum_t \langle \tilde{a}_t\tilde{a}_{t+k}\rangle}{\frac{1}{T}\sum_t \langle \tilde{a}_t^2\rangle} = \frac{\sum_t b_t\text{Var}_i(\tilde{a}^i_t)}{\sum_t \text{Var}_i(\tilde{a}^i_t)} m^k = b m^k,
\end{equation}
where we used Eq.~\eqref{eq:rkt_supp} and observe that an effective amplitude $b={\sum_t b_t\text{Var}_i(\tilde{a}^i_t)}/{\sum_t \text{Var}_i(\tilde{a}^i_t)}$ remains $k$-independent. With a $k$-independent amplitude, the second (fitting) step of the MR estimator, Eq.~(\ref{eq:step2_supp}) becomes unbiased.

For the \textit{trialseparated} method of the MR estimator~\cite{spitzner_toolbox_2018}, Eq.~\eqref{eq:rk_nonstat} translates to
\begin{equation}
    r_k = \frac{1}{N}\sum_i\frac{\left[(\tilde{a}^i_t -\left[\tilde{a}_t^i\right])(\tilde{a}^i_{t+k} -\left[\tilde{a}^i_{t+k}\right])\right]}{\left[ (\tilde{a}^i_t -\left[\tilde{a}^i_{t}\right])^2\right]},
\end{equation}
where $\left[\cdot\right]$ denotes the time average and $\left[ \tilde{a}^i_t\right]=\frac{1}{T}\sum_{t}\tilde{a}_t^i \approx 0$. Because the trials are independent but identically distributed, we can assume that $\text{Var}_t(\tilde{a}^i_t)=\left[ (\tilde{a}^i_t)^2\right]$ is constant across trials, take it out of the sum, rearrange the double sum as in Eq.~\eqref{eq:bias_sm_pre}, and find
\begin{equation}\label{eq:bias_ts}
    r_k = \frac{\frac{1}{T}\sum_t \langle \tilde{a}_t\tilde{a}_{t+k}\rangle}{\text{Var}_t(\tilde{a}^i_t)} = \frac{\sum_t b_t\text{Var}_i(\tilde{a}^i_t)}{T\text{Var}_t(\tilde{a}^i_t)} m^k = b m^k,
\end{equation}
where we used Eq.~\eqref{eq:rkt_supp} and observe that an effective amplitude $b={\sum_t b_t\text{Var}_i(\tilde{a}^i_t)}/{T\text{Var}_t(\tilde{a}^i_t)}$ remains $k$-independent. With a $k$-independent amplitude, the second (fitting) step of the MR estimator, Eq.~(\ref{eq:step2_supp}) becomes unbiased.

To summarize, we showed that the corrected time series $\tilde{a}^i_t$ enables the application of the MR estimator~\cite{wilting_inferring_2018, spitzner_toolbox_2018} for an unbiased estimation of the internal dynamics ($m$ or equivalently $\tau$) from subsampled data despite a time-dependent cyclostationary external input rate.

\begin{figure*}[t]
	\centering
	\includegraphics[width=0.8\textwidth]{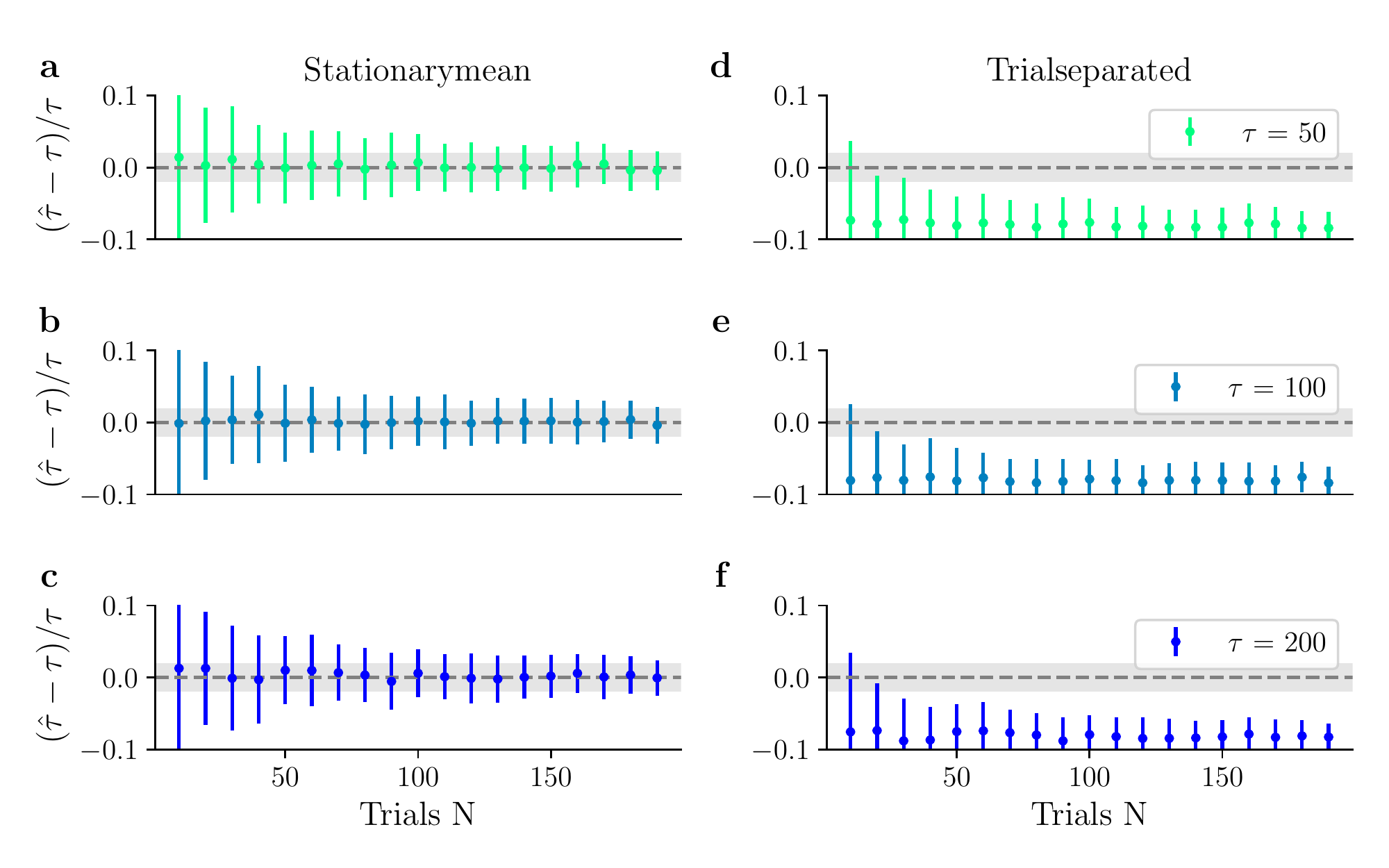}
	\caption{
		Proof of concept B. Numerical validation of the trial-ensemble average correction for subsampled BP with time-dependent external input rate (step function as in Fig. S1) for various numbers of available trials $N$ and autocorrelation times $\tau$. Error bars denote standard deviation from 100  simulations with independent estimates via MR toolbox~\cite{spitzner_toolbox_2018}. Simulation parameters (a-e): $T=100 \text{ } \tau$, $\alpha=0.01$, $\left\langle A_0 \right\rangle= 5 000$, $\left\langle h_\mathrm{up} \right\rangle / \left\langle h_\mathrm{down} \right\rangle = 2$.
		\textbf{(a-c)} Method \textit{stationarymean}. The estimates lie in the range of \SI{2}{\percent} relative deviation (grey shading) for all simulated autocorrelation times $\tau$ and subsampling fractions $\alpha$.
		\textbf{(d-f)} Method \textit{trialseparated}. The estimates suffer as expected from a short-trial bias~\cite{marriott_bias_1954, spitzner_toolbox_2018}.
		\label{fig:poc_N}
	}
\end{figure*}

\subsection{S.7 Proof of concept}
\label{sec:poc}
To verify the trial-ensemble average correction methodology, a numerical test was performed, see Fig.~\ref{fig:poc}. More specifically, branching process trials with a time-dependent external input (step function) were simulated for various autocorrelation times $\tau$ and subsampling fractions $\alpha$. Both estimation methods \textit{stationarymean} and \textit{trialseparated} of the MR estimator toolbox were tested, see Sec.~S.4. 
For \textit{stationarymean} (Fig.~\ref{fig:poc} \textbf{a-c}), the length of each trial was chosen as  $T= 100 \text{ } \tau$, where $N=5000$ trials. 
For \textit{trialseparated} (Fig.~\ref{fig:poc} \textbf{d-f}), the length of each trial was chosen as  $T= 5000 \text{ } \tau$ (to avoid short-trial biases, cf. Sec.S.4 and Refs~\cite{marriott_bias_1954, spitzner_toolbox_2018}), where $N=200$ trials were simulated. In both cases, the relative height of the step function is $\left\langle h_\mathrm{up} \right\rangle / \left\langle h_\mathrm{down} \right\rangle = 2$. In Fig.~\ref{fig:poc}, we plot independent estimates for each $\tau$ (after trial-ensemble average correction) together with the fit error from the MR estimation toolbox~\cite{spitzner_toolbox_2018}. One can clearly see that the results differ less than $1\%$ from the true $\tau$, and moreover that about $2/3$ of the results are correct within bootstrap errorbars as they should be. This demonstrates the general applicability of our method despite subsampling.

In addition, we checked the applicability of our methodology for small trial numbers $N$ (Fig.~\ref{fig:poc_N}). Again, we simulated branching processes with different autocorrelation times $\tau$ and a time-dependent external input (step function), but now we fixed the trial length $T=100\text{ }\tau$ and the subsampling fraction $\alpha=0.01$ and varied the number of trials $N$. For each data point in Fig.~\ref{fig:poc_N}, we generated 100 simulations with independent estimates of $\tau$, and we plotted the mean and standard deviation (as errorbars). For \textit{stationarymean} (Fig.~\ref{fig:poc_N} \textbf{a-c}), the standard deviation decreases with increasing trial number as expected (it starts with $10\%$ for $N=10$, which would still enable one to quantify the order of magnitude, but falls to about $2\%$ for $N\approx200$), while the mean always coincides with the true $\tau$. For \textit{trialseparated} (Fig.~\ref{fig:poc_N} \textbf{d-f}), the variance also decreases for increasing $N$, but the mean no longer coincides with the true $\tau$. This is due to the before mentioned short-trial bias~\cite{marriott_bias_1954, spitzner_toolbox_2018} given the short trial length of $T=100\text{ }\tau$. This shows that for cyclostationary external input, the best choice of methods from the two above is the \textit{stationarymean}.

\begin{figure*}[!ht]
	\centering
	\includegraphics[width=0.9\textwidth]{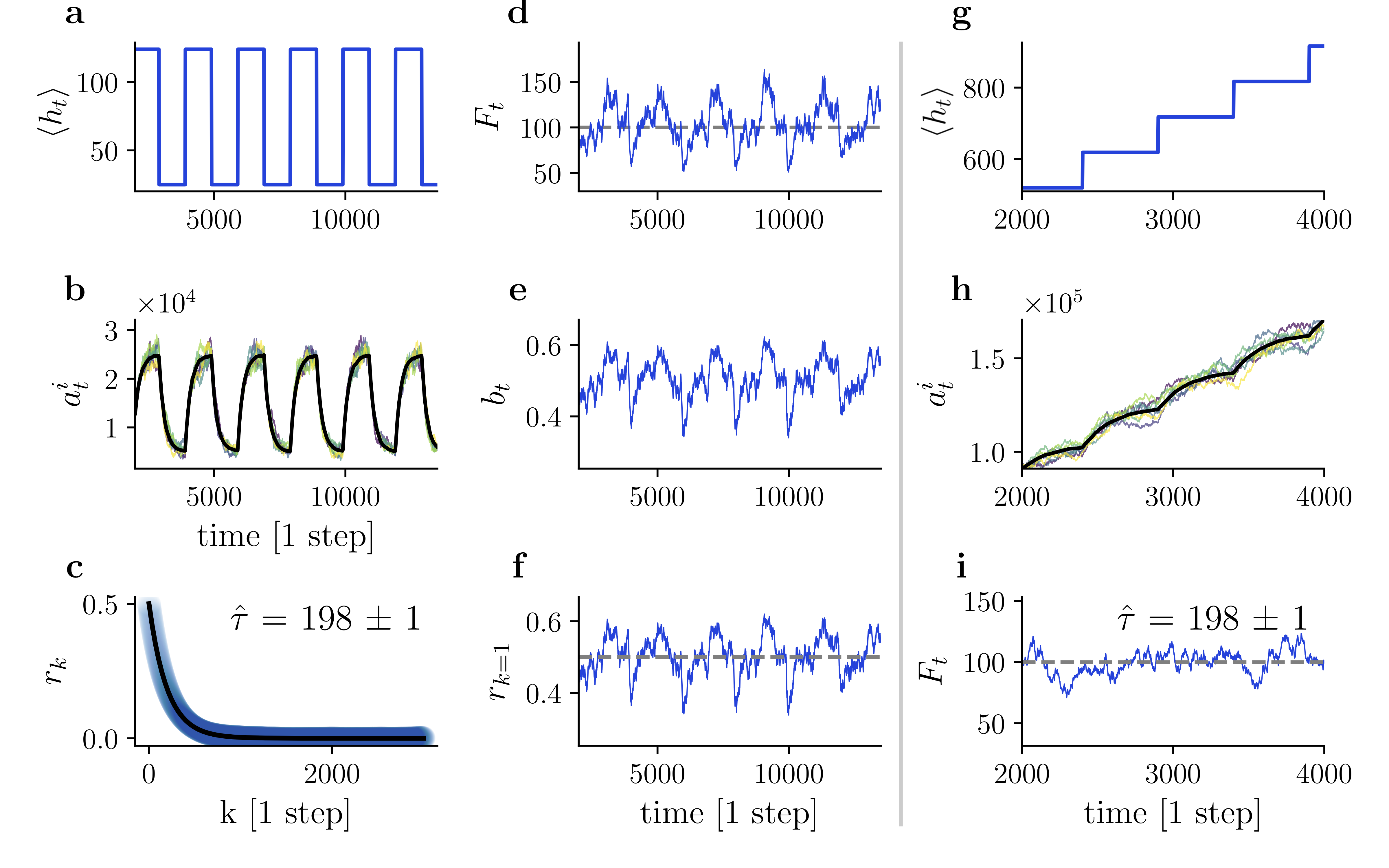}
	\caption{
	    Autocorrelation time $\tau$ can be reliably inferred even for BP with strong periodically-alternating input rate \textbf{(a-f)} and periodically-increasing input rate \textbf{(g-i)}. 
	    \textbf{(a)}~Extreme example of a periodically-alternating input rate $\left\langle h_t \right\rangle$.
	    \textbf{(b)}~Response of BP to external input with time-dependent rate (a) and internal autocorrelation time $\tau = 200$ steps, simulation length $T=1000\text{ }\tau$, subsampling fraction $\alpha = 0.01$, number of trials $N=200$, relative step height $\left\langle h_\mathrm{up} \right\rangle / \left\langle h_\mathrm{down} \right\rangle = 5$ and period of input rate $T_P=10 \text{ }\tau$. Colored lines are individual realizations, black line is trial-ensemble average.
	    \textbf{(c)}~Autocorrelation function of trial-ensemble average corrected time series, from which MR estimation yields correct internal autocorrelation time within errors.
	    \textbf{(d)}~Time-dependent Fano factor and time average (dashed line).
	    \textbf{(e)}~Time-dependent autocorrelation strength calculated from (d) via Eq.~(7).
	    \textbf{(f)}~Time-dependent leading regression slope from (e) and $m=e^{-1/\tau}\approx0.995$ together with time average (dashed line) shows that reliable estimates of the autocorrelation function can be obtained by averaging $r_k$ over time.
	    \textbf{(g)}~Extreme example of a periodically-increasing input rate $\left\langle h_t \right\rangle$.
	    \textbf{(h)}~Response of BP to external input with time-dependent rate (g) and internal autocorrelation time $\tau = 200$ steps, simulation length $T=1000\text{ }\tau$, subsampling fraction $\alpha = 0.01$, number of trials $N=200$, step height $\Delta \left\langle h \right\rangle = 100$ and period of input rate $T_P=2.5~\tau$. Colored lines are individual realizations, black line is trial-ensemble average.
	    \textbf{(i)}~Time-dependent Fano factor and time average (dashed line) of (h). The autocorrelation function is similar to (c).
		\label{fig:Fano_and_Variance_extreme_periodic}
	}
\end{figure*}

\subsection{S.8 Effect of time-dependent autocorrelation strength $b_t$}
As we derived in Eq.~\eqref{eq:b_t_supp}, regimes with rapid changes in the input rate leads to a variation of the Fano factor $F_t$ and under subsampling subsequently the amplitude $b_t$. More specifically, a rapid increase in $\left\langle h_t \right\rangle$ decreases $F_t$ and a rapid decrease in $\left\langle h_t \right\rangle$ increases $F_t$. These regimes of rapid change will be called transient regimes.  

To test the influence of strong transients on the estimation of $\hat{r}_k$ and subsequently $\tau$, a branching process of extreme transients was simulated and analyzed with the MR estimator (Fig.~\ref{fig:Fano_and_Variance_extreme_periodic}). More specifically, a periodically recurring jump in the external input rate $\left\langle h_t \right\rangle$  was implemented on a branching process with $\tau=200$ steps (Fig.~\ref{fig:Fano_and_Variance_extreme_periodic}~\textbf{a-f}). The period of the external input was chosen as $T_P=10 \text{ } \tau$ so that the up and down transients would cover five autocorrelation times each. This way approximately stationary regions were excluded and the process only consists of transient regimes. The jump's height in activity and input rate is $\left\langle h_\mathrm{up} \right\rangle / \left\langle h_\mathrm{down} \right\rangle= 5$ and the subsampling fraction $\alpha=0.01$. The estimated $\hat{\tau}$ deviates only $\SI{2}{\percent}$ from the internal autocorrelation.  The same check was repeated for a periodically-increasing input rate (Fig~\ref{fig:Fano_and_Variance_extreme_periodic}~\textbf{g-i}) with the same effect. Note that the minor systematic underestimation of the internal timescale $\hat{\tau}<\tau$ is a result from a finite-statistics bias~\cite{marriott_bias_1954}.
We conclude that the MR estimator applied to trial-ensemble average corrected time series correctly infers the internal autocorrelation time despite strong time dependence of the input rate.

\subsection{S.9 Numerical data}
\label{sec:supp_sim_and_analysis}
All branching process simulations were performed with C++, where the random number generator \textsc{MT19937} was used to drive two Poisson distributions for the branching processes recurrent internal activation and the external input. Reproducible seeding was utilized. Subsampling from the full activity was implemented numerically by drawing from a Binomial distribution.

\subsection{S.10 Epidemiological recordings}
Case report data for measles and norovirus infections in Germany were obtained from the Robert-Koch-Institute \cite{noauthor_survstatrki_nodate}. Strong seasonal fluctuation and presumably non-stationarities with a period of 52 weeks motivated the investigation of the epidemiological case reports, pre-processed with the trial-ensemble average correction and the MR estimator. The case numbers were available with a weekly binning for 52+1 weeks per year from 2001 to 2018. Week 53 was omitted due to overlapping and only full years were used, thus ignoring the first recording year. The data was separated into trials representing one year each - in agreement with the 52 week periodicity of the fluctuations.

\subsection{S.11 Spike data of \textit{macaque mulatta}}
The monkey experiments were performed according to the German Law for the Protection of Experimental Animals, and were approved by the Regierungspr\"asidium Darmstadt. The procedures also conformed to the regulations issued by the NIH and the Society for Neuroscience. 

Spike data from electrophysiological recordings in the brain  of \textit{macaque mulatta} monkeys has been analyzed. The dataset originates from a visual short-term memory experiment by Pipa et al. \cite{pipa_performance-_2009}. The monkeys were presented visual sample stimuli, which they had to remember for $\SI{3}{\second}$. Afterwards test stimuli were shown, which the monkeys had to classify into matching and non-matching.

16 single-ended micro-electrodes and tetrodes in a 4x4 grid were placed in the lateral prefrontal cortex of the trained monkeys. The inter-electrode spacing was between $\num{0.5}$ and \SI{1}{\milli\meter}. The setup allowed a simultaneous activity recording of single units and field potentials at $\SI{1}{\kilo\hertz}$, which was digitized and processed so that signal artefacts from licking and movement were rejected. The tetrode recordings were spike-sorted with the \textit{Spyke Viewer} software, whereas micro-electrode data was processed with the \textit{Smart Spike Sorter} by Nan-Hui Chen.

To convert the data into activity data, all simultaneous recordings were collapsed into one collective spike count and binned into $\Delta t = \SI{4}{\milli\second}$ time steps \cite{wilting_inferring_2018}. The \SI{4}{\milli\second} binning represents the timescale in which spikes propagate from one neuron to the other, motivated by the autoregressive process as a model for neural activity propagation. The trials within the sets had slightly varying length, so they were cut off at the end to share the length of the shortest trial within the set. No temporal alignment was performed. For the MR estimation the maximum time lag was chosen as half the minimum trial length thus $k_{max} = T_{min}/2$. This way enough data is available for each regression step $k$. For fitting, the offset-fit-function was used from the \textit{Mr. Estimator} toolbox that is $f_{\text{exp}}(x,\tau,b,c) = b \text{ } \exp(-x/\tau) + c$, see Sec.~S.4.
The maximum number of trials analyzed for each set was 300, where two sets had contained less than 300 trials, see Tab.~\ref{tab:monkey_results}.

\begin{table}[!ht]
	\centering
	\begin{tabularx}{\linewidth}{XXX}
Recording set & $\hat{\tau}$ in [ms] & \# of trials \\ 
\hline 
C001 & 88 $\pm$ 2 & 300 \\ 
C002 & 130 $\pm$ 5 & 300 \\ 
C012 & 345 $\pm$ 26 & 300 \\ 
L001 & 175 $\pm$ 11 & 300 \\ 
L008b & 239 $\pm$ 11 & 300 \\ 
L011b & 232 $\pm$ 11 & 300 \\ 
L012b & 270 $\pm$ 42 & 282 \\ 
L014b & 205 $\pm$ 81 & 299 \\ 
5115 & 214 $\pm$ 18 & 300 \\ 
5117 & 57 $\pm$ 4 & 300 \\ 
5144b & 344 $\pm$ 105 & 300 
\end{tabularx}
\caption{Detailed results of the spike data analysis of \textit{macaque mulatta}. The estimated $\hat{\tau}$ are obtained with the novel trial-ensemble average correction.
		\label{tab:monkey_results}}
\end{table}

\end{document}